\documentclass[a4,12pt]{article}
\usepackage{geometry}
\usepackage{amsmath}
\usepackage{amssymb}
\usepackage{graphicx}
\usepackage{authblk}

\begin{document}

\title {Interacting Particles in Disordered Flashing Ratchets}

\author[1]{Jim Chacko \thanks{E-mail: chacko@iopb.res.in}}

\author[2]{Goutam Tripathy \thanks{E-mail: goutam@iopb.res.in}} 

\affil[1,2]{Institute Of Physics, Sachivalaya Marg, Bhubaneswar - 751005. India.}

\maketitle

\begin{abstract}  We study the steady state properties of a system of
  particles interacting via hard core exclusion and moving in a
  discrete flashing disordered ratchet potential. Quenched disorder is
  introduced by breaking the periodicity of the ratchet potential
  through changing shape of the potential across randomly chosen but
  fixed periods. We show that the effects of quenched disorder can be
  broadly classified as strong or weak with qualitatively different
  behaviour of the steady state particle flux as a function of overall
  particle density. We further show that most of the effects including
  a density driven nonequilibrium phase transition observed can be
  understood by constructing an effective asymmetric simple exclusion
  process (ASEP) with quenched disorder in the hop rates.
\end{abstract}



\section {Introduction}

In systems that are in thermal equilibrium, the second law forbids the
existence of a net steady current. However, by continuously driving a
system out of equilibrium, a net steady state flux can be obtained.
{\it Flashing Ratchets} are an extensively studied class of systems in
which a combination of thermal noise, a periodic asymmetric potential
and an external forcing which breaks detailed balance, results in a
net directional flux \cite{reimann2002,denisov2014}. These have become
very relevant as models of active directed transport in intracellular
processes as they offer a natural mechanism for maintaining global
currents in the absence of a global driving field
\cite{julicher1997,bressloff2013}.  The ratchet mechanism has found
practical applications in areas of controlling particle motion at nano
scale \cite{nori1,nori2}, from microfluidics to nanoscale machines
\cite{hanggi2009} and controlling motion of magnetic flux quanta in
superconductors \cite{prop1,exp1a,exp1b,prop2,exp2,exp2a,exp2b} .

While bulk of the studies on flashing ratchets involve the motion of a
{\it single} particle (or a set of non-interacting ones), it has
become apparent that in many situations interactions between the
particles are relevant. This has led to many interesting collective
effects in models of elastically coupled motors
\cite{coupled1,coupled2,coupled3,reimann2002,julicher1997,bressloff2013}
and excluded volume interactions
\cite{ir1,ir2,ir3,ir4,ir5,ir6,ir7,ir8,ir9,ir10,ir11,chou2011}.  Of
particular interest to the present work is the model of a discrete
flashing ratchet considered in \cite{agha1999,menon2006} in which
large scale properties of a system of hard-core particles moving in a
ratchet potential defined on a discrete lattice was studied.

Further, it has been realized that structural non-uniformities or
bottlenecks result in frozen disorder that can occur in the tracks
along which molecular motors move \cite{chou2004}. Hence, it is
natural to study the effects of frozen or quenched randomness on the
steady state and dynamics of particles moving in flashing ratchets.
Theoretical investigations of effects of this type of disorder on
single particle motion in ratchets were reported in
\cite{harms1997,marchesoni1997}. In both these studies nontrivial
effects of disorder on mobility and diffusion were found.

In this article, we explore the effects of quenched disorder in the
ratchet potential on the steady state properties of the system of
particles interacting via  hard core constraint. We consider the
discrete ratchet model (Model I) studied in \cite{agha1999} and
introduce disorder by modifying the shape of the potential over
randomly chosen but fixed periods thereby breaking the exact
periodicity of the ratchet potential. We find surprisingly close
analogy of the disordered ratchet system with an asymmetric exclusion
process with quenched but random hop rates (DASEP)
\cite{dasep1,dasep2}. We show that the latter indeed serves as
an effective model of the former with reasonable quantitative
agreements. Although, we have studied a specific ratchet model, the
qualitative results are expected to be applicable to a whole class of
similar models.

The outline of this article is as follows. In the Section 2, we
introduce the model of the discrete flashing ratchet (FR) and discuss
construction of the equivalent asymmetric simple exclusion process
(ASEP).  Next we introduce quenched disorder in the ratchet potential
and present our results for the steady state properties of particle
flux and density profiles. We show that the results can be understood
through the equivalent disordered asymmetric exclusion process
(DASEP). We show that broadly, quenched disorder in these systems may
be classified as either {\it weak} or {\it strong} with qualitatively
distinct effects on steady state properties. In the last section, we
conclude and indicate possible future directions of study.

\section {Hard core particles moving in a discrete Flashing Ratchet (FR)}

In a discrete flashing ratchet (FR), the particle moves among a set of
discrete states, e.g., on a lattice in real space, under the influence
of a potential that flashes between an ON and an OFF state and thermal
noise.  In the present work, we consider the model introduced in
\cite{agha1999} (Model I), in which the ON state potential $V(x)$ is
defined on a one dimensional lattice of unit lattice spacing, $x=0,\pm
1,\pm 2, \cdots $. Fig.~1(a) illustrates the sawtooth potential of
integer period $w$, $V(x+w)=V(x)$ and broken reflection symmetry. The
analytical form of $V(x)$ over a period $w$ may be written as (sketched in
Fig.~1(b)) 

\begin{eqnarray}
V(x) = 
\begin{cases}
& V_0\,\,x/\mathit{a} \hspace{2.4cm}  (0\leq  x \leq  {\mathit{a}})\\ 
& V_0\,\,(w - x)/(w -\mathit{a}) \hspace{0.5cm}({\mathit{a}} \leq x \leq w),
\end{cases}
\end{eqnarray}
where $V_0$ is the peak value of the sawtooth potential and
$\mathit{a}$ is the asymmetry parameter defining the location of the
peak (for ${\mathit a}=w/2$, $V(x)$ is symmetric). We consider a
finite lattice of $L$ periods, i.e., the total number of lattice sites
in the system is $wL$. We use periodic boundary conditions to
eliminate the effects due to the edges.

\begin{figure}[h]
\centering
\includegraphics[width=0.7\textwidth]{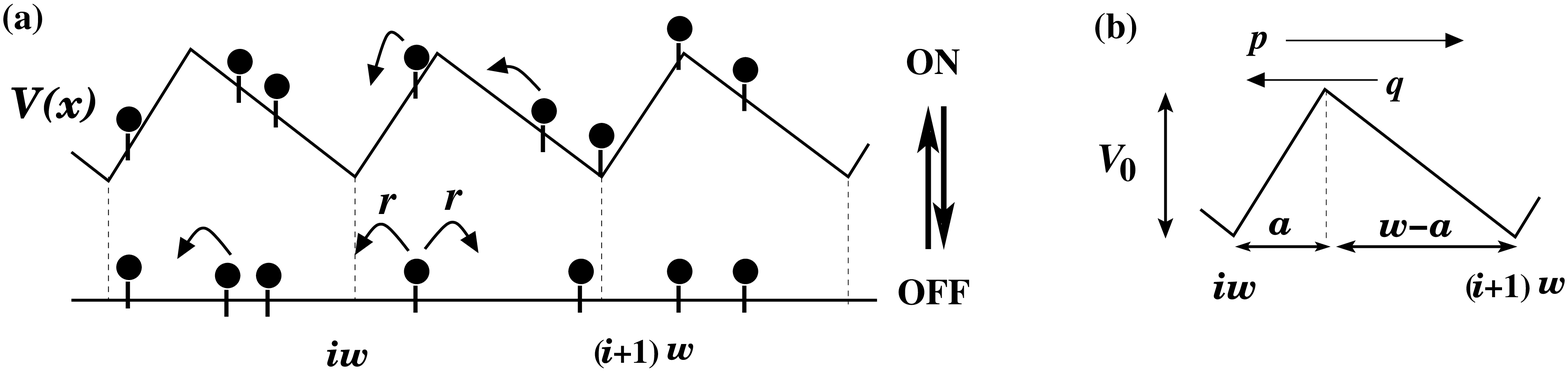}
\begin{center}
\caption{ {\bf Fig.~1(a)} is a schematic representation of the particle
  dynamics on the flashing ratchet. In the ON state, the particles
  move to neighbouring sites under the influence of $V(x)$ and thermal
  noise via hopping rates $P(n\rightarrow n\pm 1)$ (see text).  In the
  OFF state, the particles diffuse symmetrically to a neighbouring
  site with rate $r$. In both cases, attempted moves are successful
  only if the target site is empty as a consequence of the hard core
  constraint.  {\bf Fig.~1(b)} is a sketch of the $i$'th period of $w$
  sites of the sawtooth potential $V(x)$ with the asymmetry parameter
  ${\mathit{a}}$ denoting the position of the maximum $V_0$. $p,q$ are
  the effective forward and backward rates of single particle
  transitions across the $i$'th period (see text). }
\label{fig1}
\end{center}
\end{figure}

In the OFF state ($V=0$), the particles diffuse with rate $r$ to the
nearest empty neighbouring site on the left or right with equal
probability (thermal diffusion). The parameter $r$ defines the
temperature of the thermal noise ($\beta = 1/r$).  In the ON state,
particle hops between neighbouring sites $n$ and $m=n\pm 1$ satisfy
detailed balance condition with Metropolis rates: $P(n\rightarrow m) =
r* min(1,~exp[\beta(V(n) - V(m))])$.  The factor $r$ in the definition
of this rate ensures that for $V(x)=0$, the rates reduce to those in
the OFF state. This ensures that the thermal noise used in the ON
state is the same as in the OFF state. In one {\it microstep} a
randomly chosen particle attempts to hop to either of the two
neighbouring sites chosen randomly and the transition is completed
only if the target site is empty. In a system of $N$ particles
(i.e. overall particle density $\rho = N/(wL)$), $N$ such microsteps
constitute one Monte Carlo Step (MCS) which serves as the standard
unit of time in the model \footnote{By setting the units of space and
  time to unity, quantities such as density, speed, frequency, current/flux
  are dimensionless numbers.}. The flashing ratchet is realized by
switching periodically between the ON and OFF states. This switching
is independent of the state of the particles and hence it breaks the
condition for detailed balance and, together with the asymmetry of
$V(x)$, generates a net motion of the particle to the right.

For a frequency $\omega$ of flashing, in one period of
$1/\omega$ MCS, the ON and OFF states each persist for $1/2\omega$ MCS.  The smallest possible period of flashing $V(x)$
is $2~ MCS$, i.e., the largest possible frequency is
$\omega_{max} = 0.5$.  In all our simulations reported here we have
chosen $r = 1/2$ (i.e., $\beta = 2$), $w = 6, V_0=5, \omega =
0.05$. We have chosen the frequency $\omega$ such that the steady
state flux is close the maximum possible value (except for Fig.~2(b), in
which frequency is varied).  For each simulation, the system is
initially run for $10^5 - 10^6$ MCS to achieve steady state and then
for each data point (steady state flux or density profile) the
averaging is done over $10^6-10^7$ MCS. In the present study, we
switch between the ON and OFF states periodically, we expect
that most of the qualitative results would remain valid for a flashing
ratchet with more general distribution of switching time scales.

The steady state particle flux or current $J$ (defined as the mean
number of particle hops across any bond in 1 MCS in the steady state)
for this system is shown in Fig.~2(a) as a function of the overall
particle density $\rho $ for two values of the asymmetry parameter
$\mathit{a}=1,2$.  A similar plot was obtained in
\cite{agha1999,menon2006}, apart from an overall sign due to different
orientation of $V(x)$. The symmetry of the plot around $\rho=1/2$ is
due to the not so apparent particle-hole symmetry of the dynamics, as
was elucidated in \cite{agha1999}. To a very good approximation, the
plots are quadratic $J(\rho)=A_0\rho(1-\rho)$ \footnote{The most
  accurate fit to the plots of Fig.~2(a) is of the form $J(\rho)=
  A_o\rho(1-\rho)+A_1\rho^2(1-\rho)^2$ with $(A_0,A_1) = (0.0762,
  0.00683)$ and $(0.0393, -0.01530)$ for ${\mathit{a}}=1$ and $2$
  respectively. In fact, up to our numerical accuracy, there are no
  higher order corrections beyond the $A_1$ term, which is surprising
  and needs to be explored further.}, with strong resemblance to the
current-density plot for the Asymmetric Simple Exclusion Process
(ASEP) \cite{chou2011}. Indeed, the main point of this paper is an
explicit construction by which the discrete flashing ratchet (FR) is
mapped to an effective ASEP. As we will see in the next Section, this
construction can be naturally extended to the disordered versions of
the ratchet model thereby predicting qualitatively correct results
that are in fairly good quantitative agreements with direct numerical
simulations of the ratchet model.  

\begin{figure}[h]
\centering
\begin{center}
\includegraphics[width=0.4\textwidth]{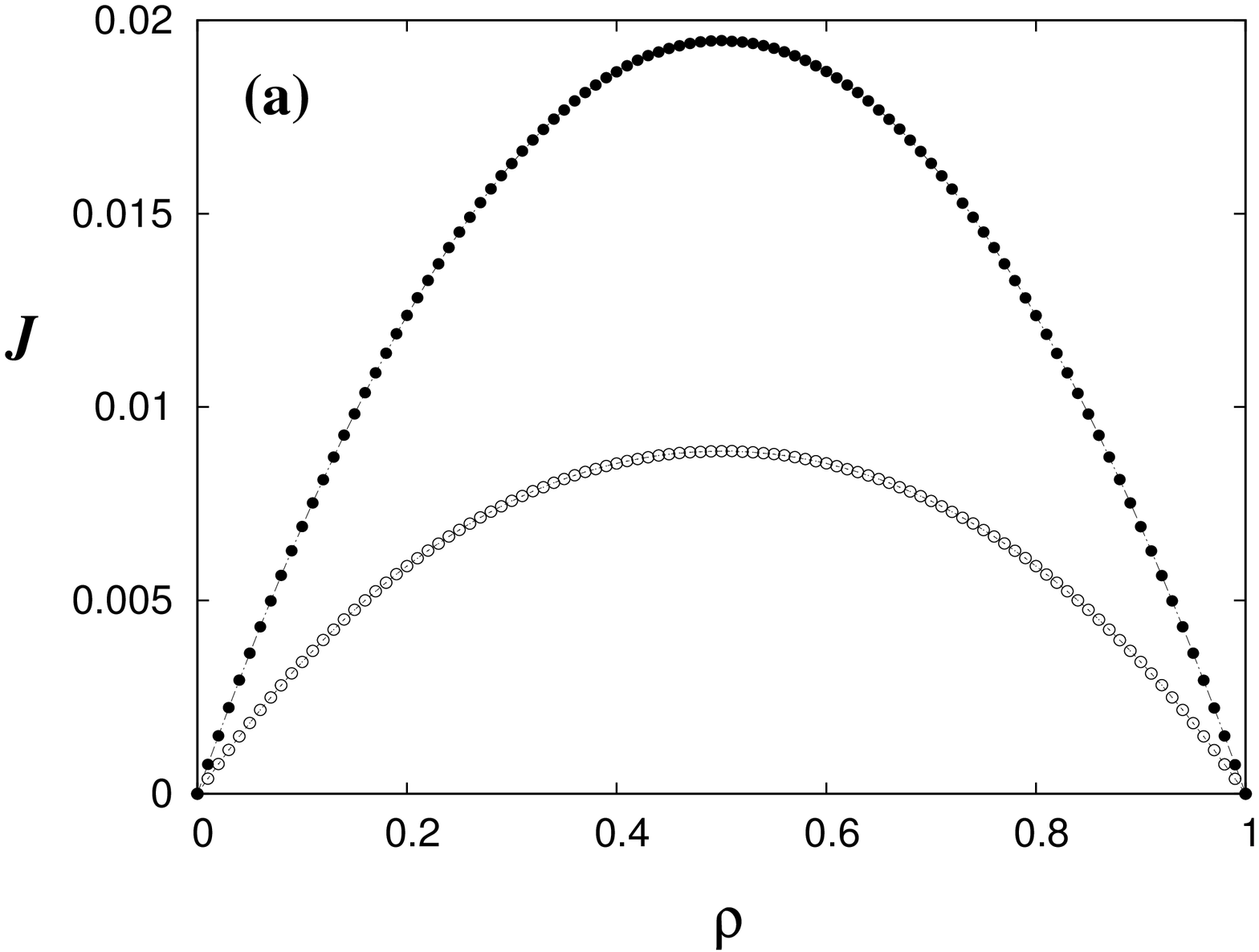}
\includegraphics[width=0.45\textwidth]{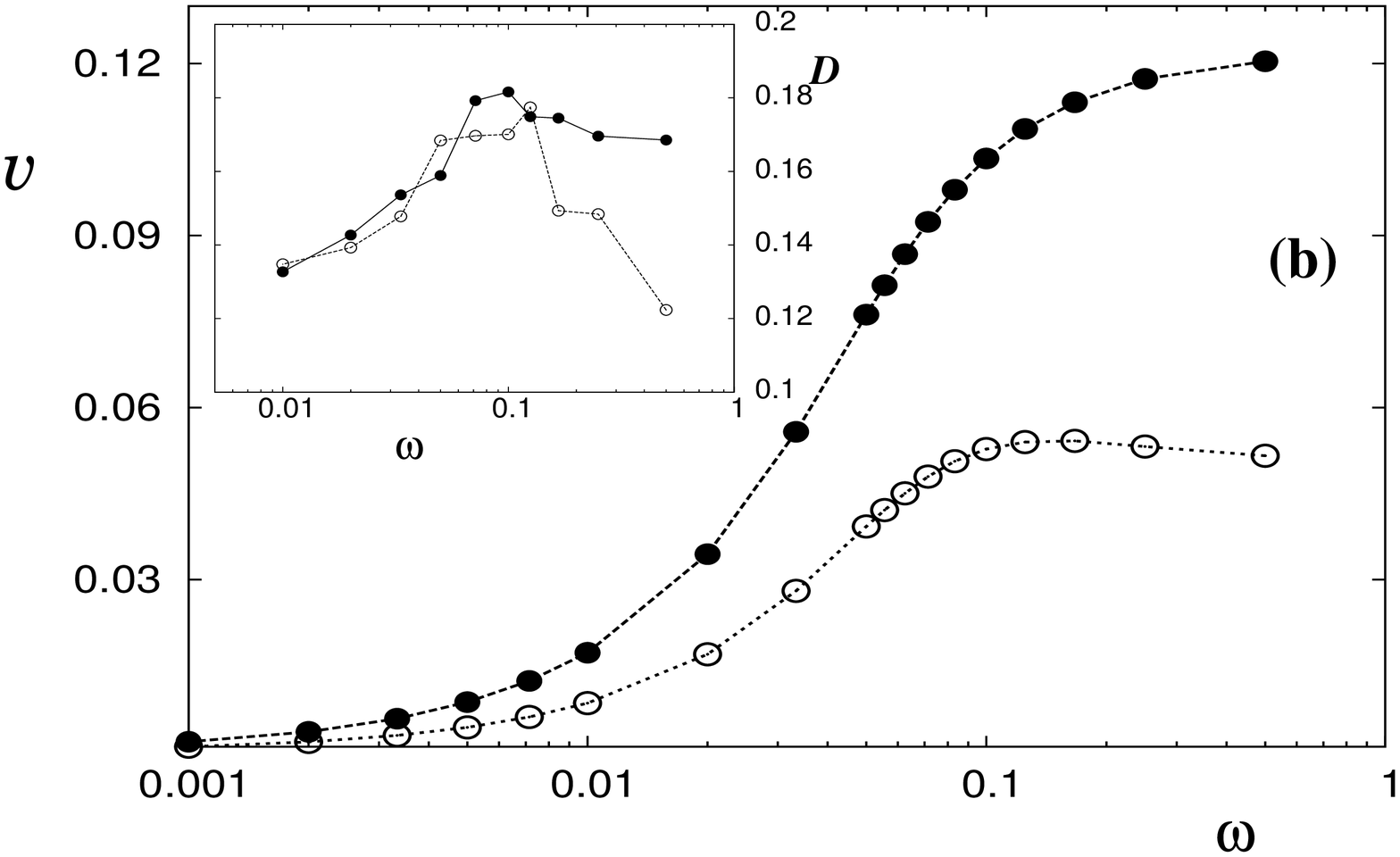}
\caption{ {\bf Fig.~2(a)} is the plot of average particle current $J$
  vs overall particle density $\rho$ for the discrete flashing ratchet
  (FR) for ${\mathit{a}}=1$ (filled circles) and $2$ (open circles)
  for $\omega=0.05$.  The plots are approximately of the form
  $J(\rho)=A_0\rho(1-\rho)$.  {\bf Fig.~2(b)} is the numerically
  computed mean velocity $v$ and diffusion coefficient $D$ (inset) for
  a single particle as a function of frequency $\omega$ of flashing in
  the FR for ${\mathit{a}}=1$ (filled circles) and ${\mathit a}=2$
  (open circles).}
\label{fig2}
\end{center}
\end{figure}

Our construction is based on the observation that the coefficient
$A_0$ measured from the $J-\rho$ graph for the interacting ratchet
matches with the mean single particle speed $v$ as shown in Fig.~2(b)
with other parameters ($V_0, r, {\mathit{a}}, w, L, \omega $)
remaining the same.  This motivates us to propose the equivalent ASEP
model as follows: It consists of $L$ sites with the $i$'th site of the
ASEP corresponding to the $i$'th {\it cell} \footnote{The $i$'th cell
  refers to sites between two maxima containing the minima at $\{iw\}$
  whereas the $i$'th period is that between the two minima $iw$ and
  $(i+1)w$. For a pure system both are equal. In the disordered
  models, all periods are of equal length $w$, cells are of unequal
  sizes. Cell average of density takes one half contribution from the
  each of the two edge sites of a cell.}  in the FR (as shown in
Figs.~3(a)(i),~3(b)(i)).  The number of particles in the ASEP equals
$L\rho$, keeping the overall particle densities the same as in the
original FR model. The following two prescriptions form the core of
this construction. (i) The transition rates across the bond $(i,i+1)$
in the equivalent ASEP, namely, $P(i\rightarrow i+1)$ and
$P(i+1\rightarrow i)$ are equal to the effective single particle
transition rates $p$ and $q$ respectively between the two cells $i$
and $i+1$ in the FR (Fig.~1(b)); and (ii) The rates $p$ and $q$
themselves depend upon the potential structure only between the minima
of the two adjacent cells ($i, i+1$) involved, i.e., the $i$'th
period. Assumption (ii) implies that $p$ and $q$ are unequal for the
FR as the potential structure encountered in the forward ($p$) and
backward ($q$) transitions are different on account of the asymmetry
of $V(x)$.  Although, both the above assumptions are strictly true at
the single particle level, and the usefulness of this construction is
validated by surprisingly close agreement between predictions based on
the equivalent ASEP model and direct numerical simulations of the pure
as well as disordered ratchet models.  The rates $p$ and $q$ are in
principle computable from the parameters of the FR
\cite{freund1999,latorre2014}, however, we compute them from the
numerically determined single particle velocity $v$ and diffusion
constant $D$ (Fig.~2(b)) using the relations $v=p-q$ and $D=(p+q)/2$
valid for a biased random walker. We adopt this computationally easier
approach for the purpose of studying the disordered systems
phenomenologically. For the two pure systems of Fig.~2(a), which we
name as FR1 (${\mathit a}=1$) and FR2 (${\mathit a = 2}$), the
numerical values of $v$ and $D$ are, respectively, $(v_1=0.076\pm
0.001,~D_1=0.15\pm0.01)$ and $(v_2=0.039\pm 0.001,~D_2=0.15\pm
0.01)$. As we will see in the next two sections, with the disordered
system being constructed from mixing these two pure systems, using
these values in the effective models leads to fairly good quantitative
agreements.

\section {Disordered (discrete) Flashing ratchet (DFR)}

We introduce quenched disorder in the flashing ratchet by breaking the
periodicity of $V(x)$ through changing the asymmetry i.e. position of
the peak ${\mathit{a}}$ in one or more periods of $V(x)$ starting from
a pure system in which all periods have ${\mathit a}=1$.  Here, we
will consider two representative cases of disordered periods:
${\mathit{a_w}}=2$ and ${\mathit{a_s}}=5$ (Fig.~3(a)). For
${\mathit{a_w}}=2$, the peak position is shifted but the sense of
asymmetry of the potential is the same as that for
${\mathit{a}}=1$. We call this {\it weak} disorder. For
${\mathit{a_s}}=5$, the asymmetry of the period is opposite to that of
${\mathit{a}}=1$. We call this {\it strong} disorder. As we will see
in the sections below, the two cases display significant qualitative
differences in the steady state properties. In the fully disordered
model a finite fraction $f$ of the $L$ periods are changed and are
distributed randomly on the lattice. We call this the Disordered
Flashing Ratchet (DFR). We also study a model in which all the
disordered periods are put consecutively in one stretch, termed as the
Fully Segregated Model (FSM) and show that a mean field approximation
becomes exact in this case. 

\begin{figure}
\centering
\begin{center}
\includegraphics[width=0.8\textwidth]{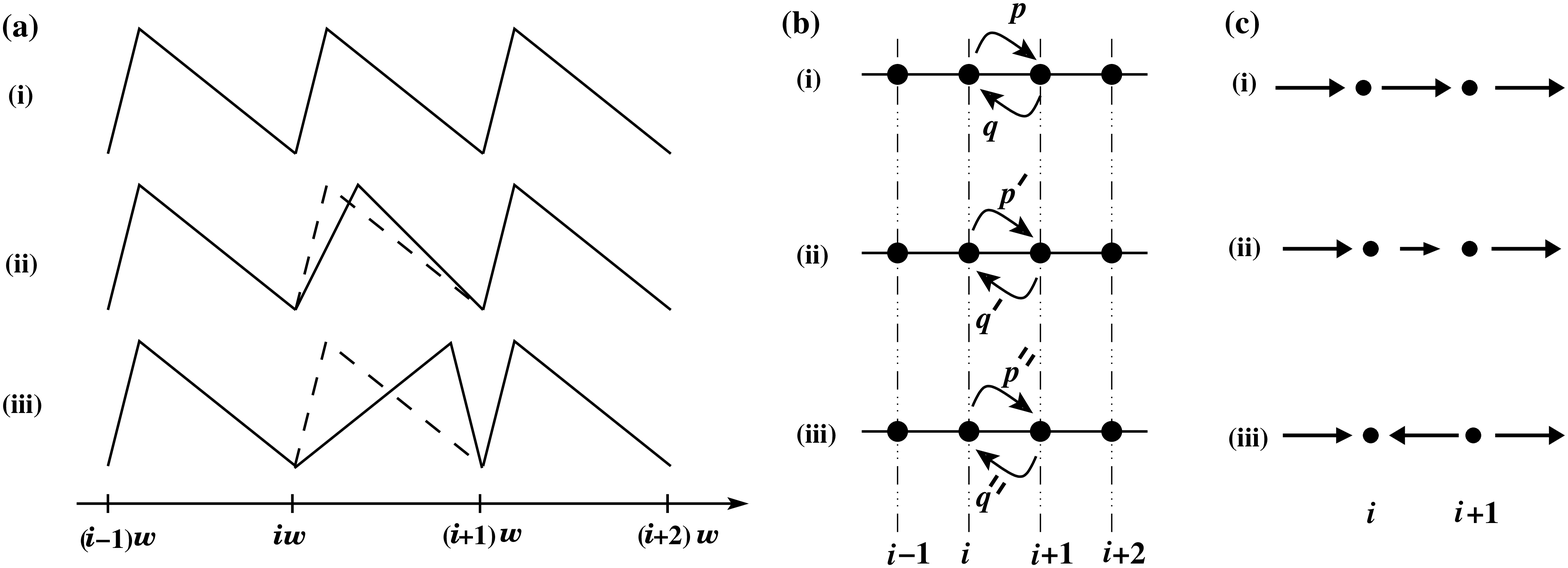}

\caption{ {\bf Fig.~3(a)} is a segment of $V(x)$ shown the (i) pure
  as well as (ii){\it weak} and (iii) {\it strong} disorder. With weak
  disorder $\mathit{a_w}=2$, the position of the peak of the sawtooth
  is changed but the sign of asymmetry remains the same. For the
  strong disorder, $\mathit{a_s}=5$, the sawtooth is flipped thus
  reversing the sign of asymmetry of that period.  {\bf Fig.~3(b)}
  are the asymmetric exclusion models corresponding to ratchet
  potentials of Fig~3a. Sites \{$i$\} of the exclusion model
  corresponds to the minima \{$iw$\} of the ratchet potential
  $V(x)$. The effective rates across the bond ($i,i+1$) corresponding
  to the disordered period are different from the rest of the
  bonds. {\bf Fig.~3(c)} is a schematic depiction of the net bias
  $p-q$ on each bond of the equivalent exclusion models of Fig.~3b. The
  direction of the arrows denote the sign of $p-q$ while the length is
  proportional to $|p-q|$. }
\label{fig3}
\end{center}
\end{figure}

\subsection {Weakly Disordered (discrete) Flashing Ratchet (wDFR)}

In Fig.~4(a), we plot $J$-vs-$\rho$ for the disordered flashing ratchet
(wDFR) for a single realization of disorder with random distribution
of two types of periods ${\mathit a}=1$ and ${\mathit a_w}=2$ (with
$f=1/2$). Although, we expect the steady state current in the DFR to
lie in between the corresponding values for the two pure systems with
$f=0$ or $f=1$ (also plotted in Fig.~4(a)), what is significant is the
appearance of a plateau for a range of densities $|\rho-1/2| \leq
\Delta$ ($\Delta\simeq 0.2$).  In this range of densities (we call it
regime $B$), the steady state flux is independent of the density and
is the maximum possible flux in the system. We call the remaining two
ranges of densities, $|\rho-1/2| > \Delta$, regime $A$.  

\begin{figure}
\centering
\includegraphics[width=0.8\textwidth]{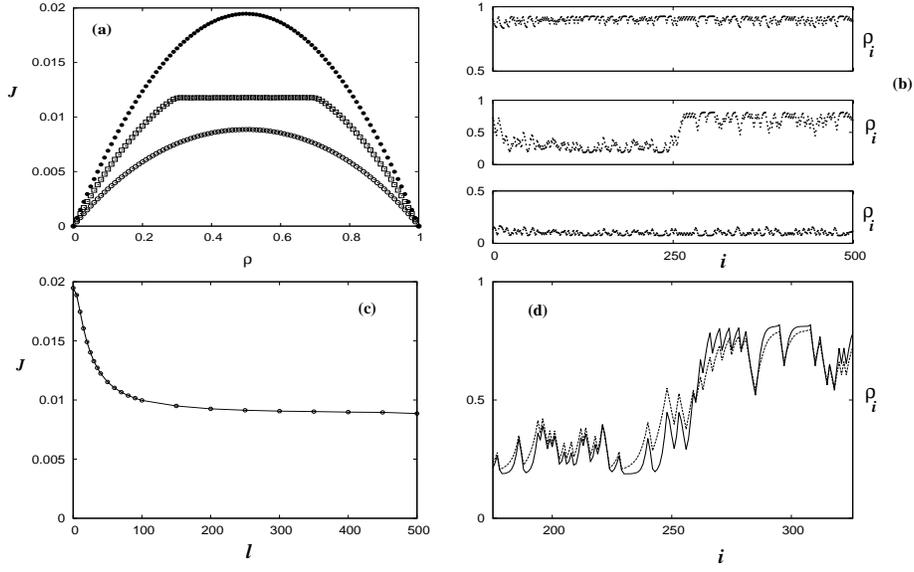}
\begin{center}
\caption
{ Steady state flux $J$ and density profiles for the wDFR. 
{\bf Fig.~4(a)} is the plot $J$-vs-$\rho$ for the weakly disordered
  flashing ratchet for $L = 500$, $f = 1/2$.  The disordered periods
  are randomly distributed on the lattice.  Also shown are the
  $J$-vs-$\rho$ for the two corresponding pure systems with
  ${\mathit{a} = 1,2}$. {\bf Fig.~4(b)} are the density profiles
  $\rho(i)$, averaged over the cell around minima $i$ of $V(x)$, for
  $\rho = 0.1$ (bottom), $0.5$ (middle) and $0.5$ (top).  
  {\bf Fig.~4(c)} is the plot of steady state flux $J$ as a function $l$
  for a system with a single stretch of disorder of length $l=fL$.
  {\bf Fig.~4(d)} is the cell-averaged density profile of the wDFR
  (solid line) for $\rho=0.5$ (middle box in (b)) compared to the
  density profile of the equivalent DASEP (dashed line). }
\label{fig4}
\end{center}
\end{figure}

Further, in Fig.~4(b), we plot the density for the wDFR averaged over
each cell as a function of the cell index $i$. Out of the three
representative density values, two values $\rho=0.1, 0.9$ (top and
bottom slides in Fig.~4(b)) belong to regime $A$ while the middle slide
(corresponding to $\rho=0.5$) belongs to regime $B$. As can be seen
from the plots, while in regime $A$ there are density
inhomogeneities on scale of a few cells, the system is homogeneous on
large length scales. By contrast, in regime $B$, there
are macroscopic regions of different densities separated by
microscopically sharp transition regions (shocks). This is akin to a
nonequilibrium phase transition which occurs as the overall density is
varied.

These two most significant features of the wDFR, namely, the shape of
the $J$-vs-$\rho$ plot and large scale density inhomogeneities in
steady state density profiles closely resembles that for the
disordered asymmetric simple exclusion process (DASEP)
\cite{dasep2}. Indeed, in the following we use the equivalent
asymmetric exclusion model constructed following the prescription of
the last section to show that the steady state results for the wDFR
found above can be qualitatively as well as semi-quantitatively
explained from the properties of the constituent pure systems.

The equivalent exclusion model for the wDFR is obtained following the
prescription of the previous section and is illustrated in
Fig.~3(c). A section of the potential with one disordered period
between minima at $iw$ and $(i+1)w$ of the FR is shown in
Fig.~3(a)(ii).  In the corresponding equivalent exclusion model shown
in Fig.~3(b)(ii), the forward ($p'$) and backward ($q'$) transition
between the corresponding sites $i$ and $i+1$ are related to $v_2$ and
$D_2$ of the FR2 through $p'={D_2+\frac{1}{2}v_2}$ and
$q'={D_2-\frac{1}{2}v_2}$ as explained in the previous section. Using
the values $(p,q)$ and $(p',q')$ for bonds in the equivalent model
corresponding to FR1 and FR2 periods respectively, in Fig.~4(d), we
compare the steady state density profiles (averaged over each cell)
for the wDFR and the equivalent DASEP for $\rho=0.5$. As can be seen
from the plot, the locations of the shocks in the wDFR and the
equivalent DASEP match exactly. The shape of the microscopic shocks
and, more remarkably, the transition region between the high density
and low density phases are captured fairly accurately by the
equivalent DASEP.

The plateau in the $J$-vs-$\rho$ plot (Fig.~4(a)) and the density
inhomogeneities (Figs.~4(b), 4(d)) for the wDFR can be qualitatively
understood in terms of the equivalent DASEP, as argued in
\cite{dasep2}. However, in the following subsection we introduce below
the fully segregated model (FSM) which displays similar phenomena as
the wDFR and at the same time is amenable to exact analysis.

\subsubsection {Fully Segregated Model (FSM)}

The Fully Segregated Model (FSM) is obtained from the wDFR by placing
the periods with same value of ${\mathit a}$ consecutively
together. Let $X$ and $Y$ denote the stretches of periods with
${\mathit a}=1$ and ${\mathit a}=2$ respectively.  This minimizes the
effects due to the small size of the stretches in wDFR. In
Fig.~5(a), we plot the $J$-vs-$\rho$ for the FSM.  We see a plateau in
the plot that is similar to that in Fig.~4(a) for the wDFR with the
quantitative difference that the peak (plateau) current now matches
with the peak current for the pure system FR2. In Fig.~5(b), we plot the
cell averaged density profiles for the FSM for a density $\rho=0.5$ in
the plateau (regime B) and two densities ($\rho=0.1, 0.9$) in regime
A. 

\begin{figure}
\centering
\includegraphics[width=0.4\textwidth]{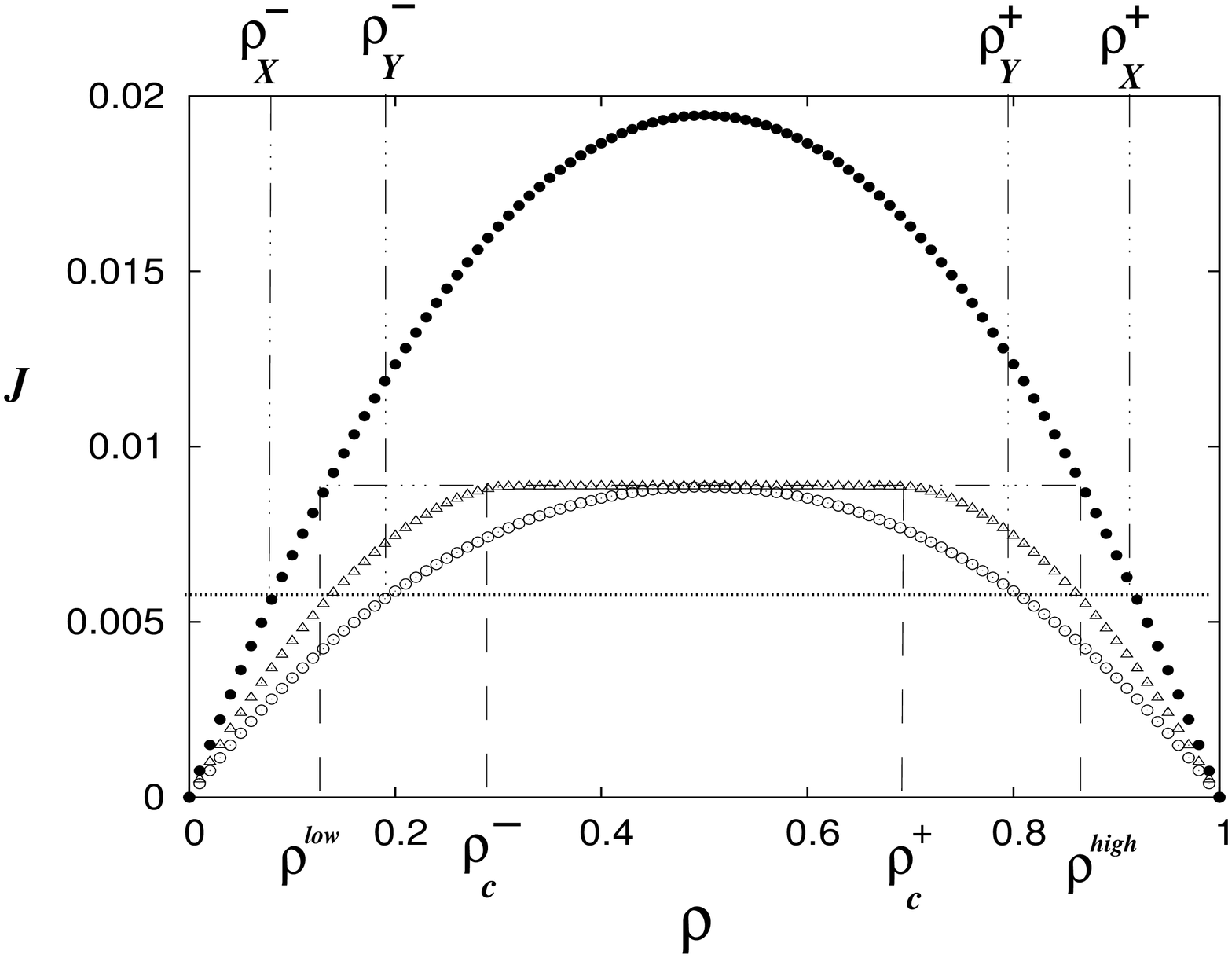}
\includegraphics[width=0.4\textwidth,height=0.28\textwidth]{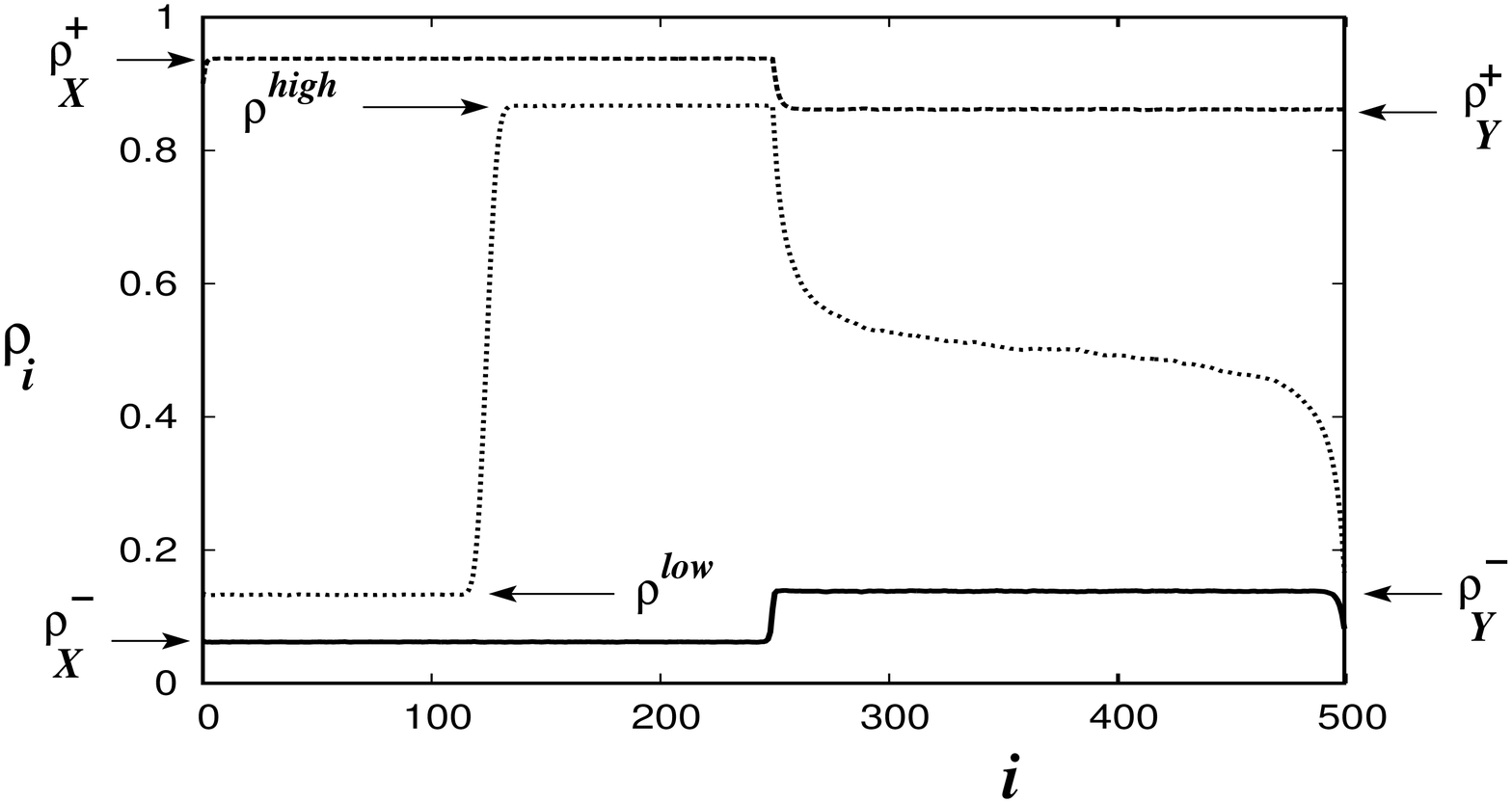}
\begin{center}
\caption
{ {\bf Fig.~5(a)} is the $J$-vs-$\rho$ plot for the FSM for the wDFR
  on a system of $L=500$.  The disordered periods are in a single
  stretch of $L/2$ periods (i.e. $f=1/2$). Also shown are the $
  J-vs-\rho $ of the two corresponding pure systems with ${\mathit{a}}
  = 1$ (filled circles), ${\mathit{a}}=2$ (open circles).  
  {\bf Fig.~5(b)} is the density profile averaged over each cell for
  different density of particles, $\rho = 0.1, 0.5, 0.9$. The $Y$
  stretch corresponds to $250<i\le 500$.}
\label{fig5}
\end{center}
\end{figure}

Let $\rho_X$ and $\rho_Y$ denote the steady state densities in the $X$
and $Y$ stretches respectively and $J$, the steady state current in
the system. In regime $A$, the cell averaged density in each segment
is uniform and hence particle conservation implies: $(1-f)\rho_X +
f\rho_Y = \rho$ and the equality of the steady state current in both
stretches implies: $A^X_0\rho_X(1-\rho_X) = A^Y_0\rho_Y(1-\rho_Y) = J$ (where
$A^X_0=v_1$ and $A^Y_0=v_2$). Solving for $\rho_Y$ gives

\begin{eqnarray}
\nonumber
\rho_Y^\pm &=& \frac{A^X_0f(f+2\rho-1) - A^Y_0(1-f)^2}{2(A^X_0f^2 - A^Y_0(1-f)^2)}  \pm \\
& & \hspace{-6ex} \frac{\sqrt{(A^X_0f(f+2\rho-1) - A^Y_0(1-f))^2 + 4A^X_0\rho(\rho+f-1)(A^X_0f^2 - A^Y_0(1-f)^2)}}{2(A^X_0f^2 - A^Y_0(1-f)^2)}, \\
& & \qquad\qquad \rho_X^\pm = \frac{\rho-f\rho_Y^\pm}{1-\rho},
\qquad\qquad J = A^X_0 \rho_X^\pm (1-\rho_X^\pm).
\end{eqnarray}
As $\rho$ is increased from $0$ (decreased from $1$), the mean density
$\rho_X$ and $\rho_Y$ of each stretch increases (decreases)
(Fig.~5(a)). Concurrently, the average current $J$ also increases until
the current in the $Y$ stretch reaches its maximum possible value
$J_Y^{max} = A^Y_0/4$; at this point, the mean density of particles in
the $Y$ stretch is $\rho_Y = 1/2$ (the deviation from a uniform value
of $1/2$ is due to the boundaries and this becomes negligible in the
limit of large value of $Y$ stretch length $(1-f)L$). This marks the
onset of regime $B$ which occurs when the overall particle density
reaches the critical value $\rho = \rho_c^-$ $(\rho_c^+)$; where
$\rho_c^\pm=\frac{1}{2}\pm\frac{1-f}{2}\sqrt{1-\frac{A^Y_0}{A^X_0}}$.
Thus, the width of the $B$ regime is $2\Delta =
(1-f)\sqrt{1-\frac{A^Y_0}{A^X_0}} =(1-f)\sqrt{1-v_2/v_1}$.  Further
increase (decrease) of the density does not affect the density in the
$Y$ stretch, rather, the excess (deficit) density is incorporated in
the $X$ stretch by the creation of a high (low) density region of
density $\rho_X=\rho^{high} (\rho^{low})$. Thus, in regime $B$,
two phases of densities $\rho^{high}$ and $\rho^{low}$ coexist
(Fig.~5(b)). The two densities, $\rho^{high}$ and $\rho^{low}$ are
related by $\rho^{high} + \rho^{low} = 1$, so that the current in
the two phases of the $X$ stretch are equal. For this current to be
equal to that in the $Y$ stretch demands
$A^X_0\rho^{high}(1-\rho^{high}) = A^X_0\rho^{low}(1-\rho^{low})
= A^Y_0/4$.  Thus, $\rho^{high,low} = \frac{1}{2}(1\pm\sqrt{1-\frac{A^Y_0}{A^X_0}})$ and the fraction of these
phases is given by (using a lever rule) $\phi^{low,high}_X =
|\rho-\rho^\pm_c|/(\rho^{high} - \rho^{low})$.

The fact that the system chooses to go in to a state with large scale
density inhomogeneity is a consequence of the the extremum current
principle that seems to be valid in regime $B$: If there are
more that one apparent choice of steady states, the system would
select the one that extremizes the mean current.  Such a principle
has been shown to be valid in ASEP with open boundaries
\cite{krug1991,popkov1999}, in disordered ASEP \cite{dasep2} and has been
more recently applied in determining steady states in molecular motor
systems \cite{ecp1,ecp2}.

Now, let us try to understand qualitatively the results obtained for
the wDFR (Fig.~4) in terms of those of the FSM.  The wDFR may be
looked upon as obtained from FSM by randomly scrambling the $X$ and
$Y$ stretches of the latter. Thereby, the density profiles
corresponding to regime $A$ of the wDFR (top and bottom slides in
Fig.~4(b)) are similarly scrambled versions of the corresponding density
profiles of the FSM (top and bottom graphs in Fig.~5(b)). The
microscopic shocks in the former are essentially the alternation
between $\rho_X^+$ and $\rho_Y^+$ (for $\rho>\rho_c^+$) or between
$\rho_X^-$ and $\rho_Y^-$ (for $\rho<\rho_c^-$). As expected, the exact
numerical values are slightly different from that of the FSM on
account of the smallness of the stretches in the wDFR.  Similarly, the
phase segregated density profile in regime $B$ of the wDFR (middle
slide in Fig.~4(b)), corresponds to an alternation of densities between
$\rho^{high}$ and $1/2$ (the {\it high} density phase) and that
between $\rho^{low}$ and $1/2$ (the {\it low} density phase). Again
the actual values of the densities are different on account of the
smallness of the stretches in wDFR. Also, density profiles within the
same stretch are far from being spatially uniform, unlike that in the FSM,
due to strong boundary induced correlation effects. 

The $J-vs-\rho$ plot for the wDFR (Fig.~4(a)) is a direct analogue of
the same plot for the FSM with the important difference being the high
value of the peak (plateau) current in the former. This is in fact a
direct consequence of the small lengths of the $Y$ stretches in the
wDFR. In Fig.~4(c), we have plotted the peak current of the FSM as a
function of the length of the $Y$ stretch.  The peak current
approaches the peak current of the FR2 beyond a $Y$ stretch length of
about $l \simeq 100$. By contrast, the length of the longest $Y$
stretch in the disorder realization of Fig.~4(a) is $l_{max} \simeq
35$. This corresponds to a peak current for the corresponding FSM of
$0.012$ which is very close to what is observed in Fig.4(a). It should
be noted that, in the thermodynamic limit (i.e., $L\rightarrow\infty$),
the largest stretch of $Y$ also becomes large ($l_{max}\sim \ln
L$) and thus the $J$-vs-$\rho$ plot
for the fully disordered wDFR coincides with that of the FSM (with the
same value of $f$). This feature of nonzero peak current in the
thermodynamic limit is what distinguishes wDFR from sDFR. As we will
show in the next subsection, in the latter, the steady state current
vanishes in the thermodynamic limit $L\rightarrow\infty$.

\subsection {Strongly  Disordered Flashing Ratchet (sDFR)}

A strongly disordered flashing ratchet (sDFR) is obtained if the
asymmetry of the disordered period is opposite to that of the other
periods. This is achieved by shifting the peak of $V(x)$ to
${\mathit{a}_s} > w/2$ while the starting model has all periods with
${\mathit a} <w/2$. We chose the sDFR model with a fraction $f$ of
periods with ${\mathit a_s}=w-{\mathit a}=5$ and a fraction $1-f$ of
periods with ${\mathit a}=1$. Clearly, the currents for the two pure
systems $f=0$ and $f=1$, which are mirror reflections of each other,
are related by $J_{w-\mathit{a}} = -\,J_{\mathit{a}}$ and therefore
for $f=1/2$, the steady state current in the sDFR vanishes. 
 
\begin{figure}
\centering
\includegraphics[width=0.4\textwidth]{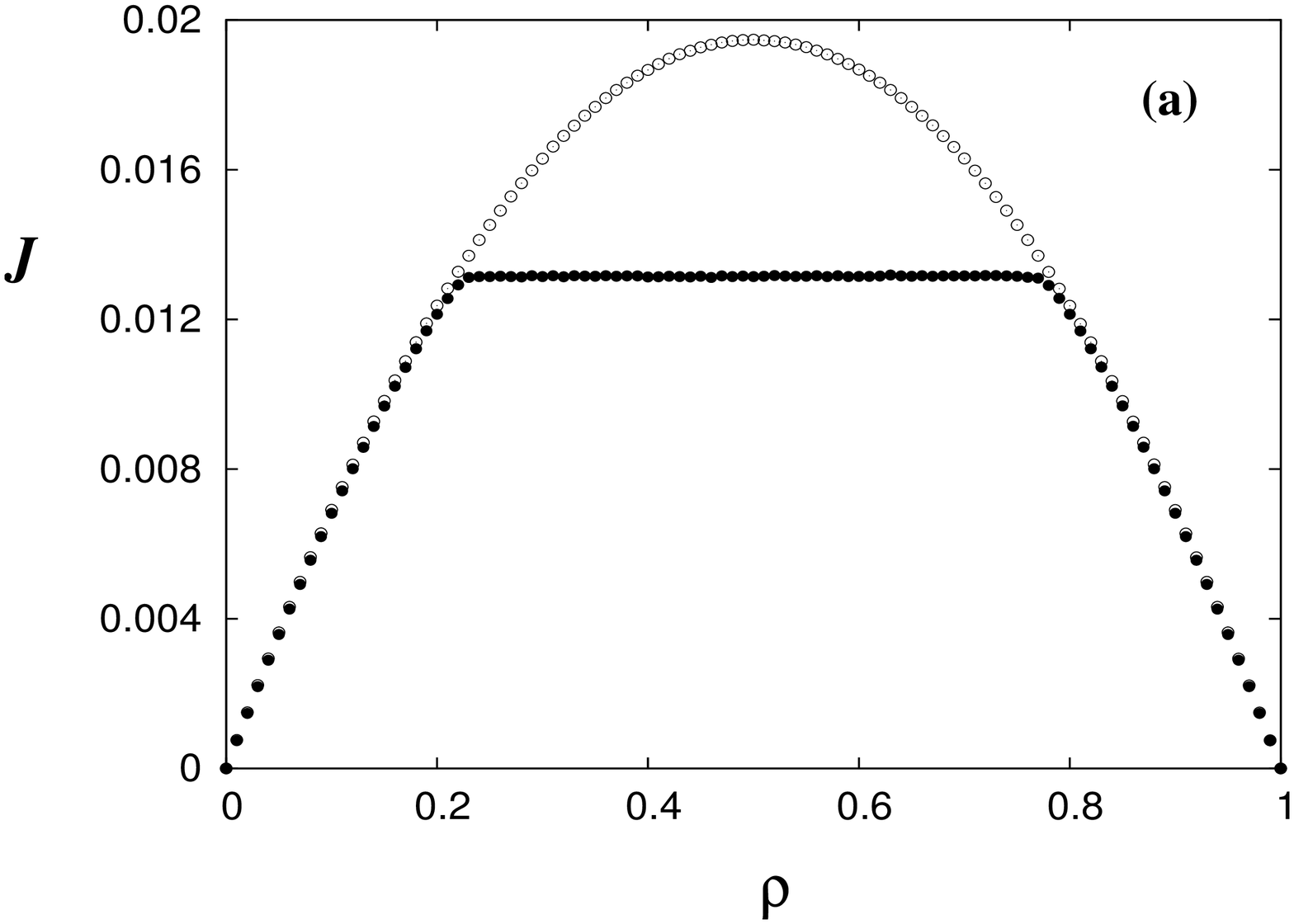}
\includegraphics[width=0.4\textwidth]{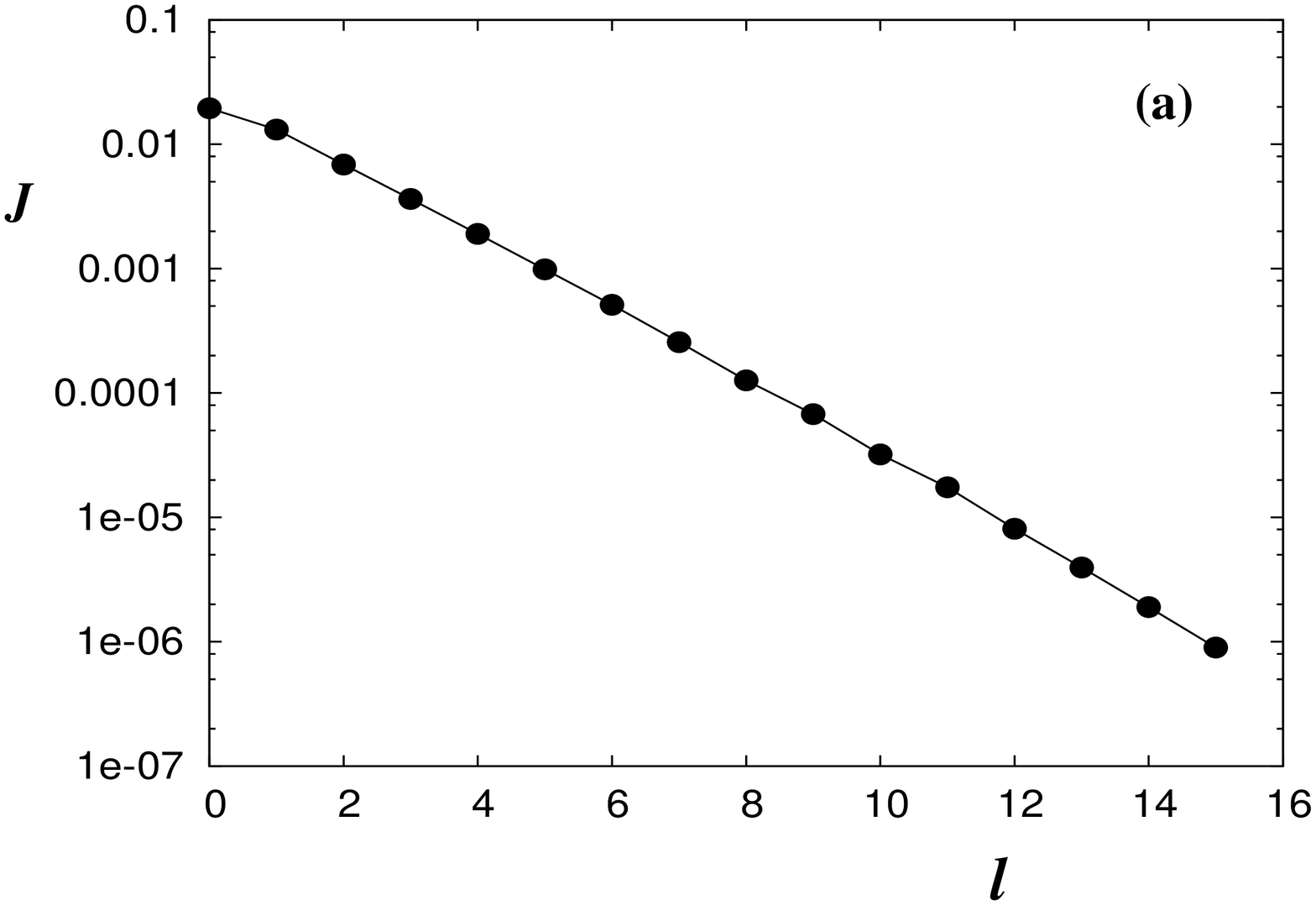}
\begin{center}
\caption
{ {\bf Fig.~6(a)} is the $J$-vs-$\rho$ plot for the for the flashing
  ratchet with strong disorder with one disorder period $f = 1/L$ for
  a system of $L=500$ periods.
  {\bf Fig.~6(b)} is  the steady state flux $J$ for
  $\rho=1/2$ with one stretch of $l=fL$ reversed periods. The flux
  vanishes exponentially with the length of the stretch.}
\label{fig6}
\end{center}
\end{figure}

In Fig.~6(a), we plot the numerically obtained $J$-vs-$\rho$ for the sDFR
for a single defect period. It is seen the same plateau structure
appears here as in the case of the wDFR, however, it should be noted
that the reduction in the peak value of the current is significant
considering there is only one defect period. As in case of the wDFR,
the peak flux is determined by the longest stretch of
disorder. Indeed, in Fig.~6(b), we have plotted the peak current as a
function of the length $l=fL$ of a single defect stretch (i.e., in the
corresponding FSM) for a system of size $L=500$ and density
$\rho=0.5$. We see that the peak current vanishes exponentially ($J_l
\sim e^{-\alpha l}$) with $l$ unlike in the case of wDFR (Fig.~4(c)),
where it approaches a finite value.  This result has significant
implication for the thermodynamic limit of the fully disordered
sDFR. In a system of $L$ periods with a finite fraction $f$ of
reversed periods, statistically, the length of the longest reversed
stretch $l_{max}$ increases a $\ln L$ \cite{barma1993,dasep2}. Since
this longest stretch of reversed periods determines the peak current
$J_{max} \sim e^{-\alpha l_{max}}$, thus, for large $L$ the peak
current $J_{max} \sim L^{-\alpha}$. I.e., in the thermodynamic limit
the flux vanishes for the sDFR. This is in contrast to the single
particle result obtained in \cite{harms1997} where the single particle
velocity vanishes only in certain ranges of flashing frequency $\omega $.

The results for the sDFR are similar to those for the DASEP with
backbends studied in \cite{dasep2}. In fact, using the construction
proposed in Section 2, the sDFR translates to a DASEP with local
reverse bias in the bonds corresponding to the disordered periods
(Fig.~3(c)). The stretches of disordered periods act as backbends and
this leads to the vanishing of the particle flux in the thermodynamic
limit \cite{barma1993}.

\section {Results and Discussions}

In this article we have studied the effect of quenched disorder in a
one dimensional discrete model of the flashing ratchet with hard core
particles (motors).  We measure the steady state particle flux as
density of particles and disorder are varied. We have shown that
disorder can be classified broadly as strong or weak depending upon
whether the steady state flux vanishes or not in the thermodynamic
limit. While in case of weak disorder, the steady state flux has a non
zero finite value in this limit, that in case of strong disorder
vanishes as a power law of system size.

We further show that, in disordered ratchets, the fundamental
diagram ($J$-vs-$\rho$ plot) has a plateau, i.e., the maximum particle
flux is independent of the overall particle density for a range of
densities.  Inspection of the spatial density profile shows that
corresponding to this plateau regime there is large scale spatial
segregation of high and low density phases in the disordered
system. Through a mapping to an equivalent asymmetric exclusion model
using single particle parameters of the ratchet, we show that these
features for the disordered interacting ratchets are qualitatively
similar as well as quantitatively close to those observed for the
disordered asymmetric exclusion process (DASEP).

\section{Conclusions}

As a test of the basic prescription, proposed in the present report,
of extending single particle rates and the assumption that these rates
can be locally determined needs to be checked for different forms of
disordered ratchets, e.g., by introducing more than two types of
periods as has been considered here, or introducing disorder via
having periods of different lengths (i.e., disorder in $w$) and/or in
the potential barrier height $V_0$ in the ON state. Further, effects of
disorder in systems with open boundaries is a possible area to
explore.

In this paper we have focused our attention on the steady state
properties of the systems. The dynamics of fluctuations in the steady
state is expected to show interesting behaviour considering the
results for the same in the disordered ASEP. E.g., as noted in
\cite{agha1999}, in the phase segregated state, the vanishing of the
kinematic wave speed may lead to a different dynamic universality
class for the fluctuations. This would be an interesting aspect to
explore.

\end{document}